\documentclass[aps]{revtex4-1}
\usepackage{graphicx}
\begin{document}
\title{Spin-Transfer-Torque Driven Magneto-Logic OR, AND and NOT Gates}
\author{C. Sanid and S. Murugesh}
\email{murugesh@iist.ac.in}
\affiliation{Department of Physics, Indian Institute of Space Science and Technology, Trivandrum 695547, India}
\begin{abstract}
{We show that current induced magneto-logic gates like AND, OR and NOT can be designed with the simple architecture involving a single nano spin-valve pillar, as an extension of our recent work on spin-torque-driven magneto-logic universal gates, NAND and NOR. Here the logical operation is induced by spin-polarized currents which also form the logical inputs. The operation is facilitated by the simultaneous presence of a constant {\it controlling} magnetic field, in the absence of which the same element operates as a magnetoresistive memory element. 
We construct the relevant phase space diagrams for the free layer magnetization dynamics in the monodomain approximation and show the rationale and  functioning of the proposed gates. The flipping time for the logical states of these non-universal gates is estimated to be within nano seconds, just like their universal counter parts.
} 
\end{abstract}

\maketitle
\section{Introduction}
The phenomenon of Spin-Transfer-Torque (STT) predicted independently by Slonczewski\cite{slonc:1996} and Berger\cite{berger:1996} in the nineties has been a turning point in the inception and development of many current controlled magnetic multilayer devices like Magnetic Random Access Memories (MRAMs), Magneto-Logic devices and the Spin-Torque Nano Oscillators (STNOs). At the heart of these Spintronic devices is the phenomenon of Giant Magneto Resistance (GMR) in conjunction with STT effect, which help realizing many of its potentialities. The STT effect involves the reorientation of magnetization of a ferromagnetic thin film due to a spin polarized in either a current-in-plane (CIP) geometry or perpendicular to it's plane (CPP) geometry. Out of many theoretically and experimentally studied aspects of STT, the current induced magnetization switching and self-sustained microwave oscillations are of special importance to the spintronics industry \cite{stiles:2006,wolf:2006,myers:1999,gro:2001,kiselev:2003,rippard:2004,berkov:2008}.

The aspect of {\it non-volatility}, fundamentally inherent in the system, and the significant reduction in power consumption have prompted the development of spin-valves as memory devices. The earlier proposals, however, were based on a field induced 
magnetic switching(FIMS) approach for writing data, which uses two orthogonal pulses of magnetic filed to achieve writing. Magnetic random access memory\,(MRAM) models based on current induced magnetic switching (CIMS), wherein STT phenomenon forms the core, have since been proposed. Apart from the more obvious application as plain memory storage devices, spin valve based magneto-logic devices have also been attempted in the recent past. FIMS based field programmable logic gates using GMR elements were proposed
by Hassoun {\it et al.}\cite{black:1997}, wherein the type of the logical operation to be performed can be altered by additional fields. Further models have also been suggested where the logical state of the GMR 
unit is manipulated using FIMS \cite{richter:2002,koch:2003,ney:2005,wang:2005,chao:2007}. Similar programmable models 
based on spin valve magneto-logic devices are also known in literature\cite{zhao:2006,dery:2007,buford:2011}. 
These later models, based on CIMS, involve additional spin-valve elements that together form a single logical unit, or more than one current carrying plate capable of generating fields in orthogonal directions. Besides, in these models, bi-polar
currents were crucial in writing or manipulating data. Invariably, this requires a more complex architecture than is required for a simple magnetic memory unit. Recently we have proposed CIMS based universal magneto-logic gates, NOR and NAND, using spin-valve pillars, wherein a magnetic field is held constant throughout the logic operation, and acts as a control field\cite{sanid:2012}. In the absence of the magnetic field, the same spin valve element functions as a simple magnetic memory element, written by a spin-current. The major advantages of our model were the simplicity of architecture requiring a single spin-valve pillar for fabrication and the usage of uni-polar currents and fields for any particualr universal gate. 

In this paper we extend our work on alternative magneto-logic gate models, to focus on logic OR, AND and NOT gates. The previous work had to invoke a re-interpretation of logic $1$ and logic $0$ in order to develop the NAND gate. However, we see that all the non-universal gates can be designed with a consistent interpretation of logical states. In these proposed models, as with the NAND and NOR gates, we use STT for writing, while the magnetic field is held constant in magnitude and is required only during the logical operation. Thus the applied field acts as a control switch for the gates. The gates are also non-volatile, as naturally expected in a magneto-logic device. 

\section{Spin-Valve Pillar Geometry and the Governing Landau-Lifshitz-Gilbert 
Equation}

The system under consideration is a regular spin valve primarily consisting of a conducting layer sandwiched between two ferromagnetic layers, one pinned and the other free, with the magnetization in the pinned layer parallel to the plane
of the free layer - a geometry that is well studied\cite{kiselev:2003,albert:2000,mangin:2006}. Further, the free layer is also subject to a constant Oersted field, by a conducting plate carrying current. The dynamics of the macrospin magnetization of the free layer is governed by the Landau-Lifshitz-Gilbert\,(LLG) equation with the STT term, whose dimensionless form 
is given by\cite{berkov:2008,bert:2005}
\begin{equation}\label{llg}
\frac{\partial \textbf{m}}{\partial t}-\alpha\textbf{m}\times\frac{\partial \textbf{m}}{\partial t} = - \textbf{m} \times \textbf{H}_{eff},
\end{equation}
where
\[
\textbf{H}_{eff} = \left( \textbf{h}_{eff}-\beta\frac{\textbf{m}\times\textbf{e}_{p}}{1+c_{p}\textbf{m}\cdot\textbf{e}_{p}}\right).
\]
The free-layer magnetization $\textbf{m}$ and the effective field $\textbf{h}_{eff}$ are normalized by the saturation magnetization $M_s$. Time is measured in units of $(\gamma M_s)^{-1}$, where $\gamma$ is the gyromagnetic ratio (for Co layers, for instance, this implies
time scales in the order of picoseconds). The constant $\alpha$ is the damping factor and unit vector $\textbf{e}_{p}$ is the direction of pinning ($\hat{x}$ in our case, and in plane). The other constant $c_{p}\,({1}/{3}\le c_{p}\le1)$ is a function of degree of spin polarization $P\,(0\le P\le1)$:
\begin{equation}\label{cp}
  c_{p}=\frac{(1+P)^{3}}{3(1+P)^{3}-16P^{{3}/{2}}}
\end{equation}
In the numerical calculations that follow we have used the typical value of $P=0.3$. The parameter $\beta$ is proportional to the spin current density (typically of the order of $10^{-2}$ for Co layers, with current densities $\sim 10^{8}$~A/cm$^2$). 
The effective field is given by 
\[
\textbf{h}_{eff} = h_{ax}\hat{x}-(D_{x}m_{x}\hat{x}+D_{y}m_{y}\hat{y}+D_{z}m_{z}\hat{z}),
\]
where $h_{ax}\hat{x}$ is the external field and $D_{i}$s$(i=x,y,z)$ are constants that reflect the crystal shape and anisotropy effects. Particularly, we chose our film such that the anisotropy is in-plane, and also lies along the $x$-axis. The plane of the free layer is chosen to be the $x-y$ plane. With this choice $D_is$ are such that $D_{x}<D_{y}<D_{z}$, making $\hat{x}$ the free-layer easy axis.

\begin{flushleft}
\begin{figure}[h]\label{para}
\includegraphics[width=1\linewidth]{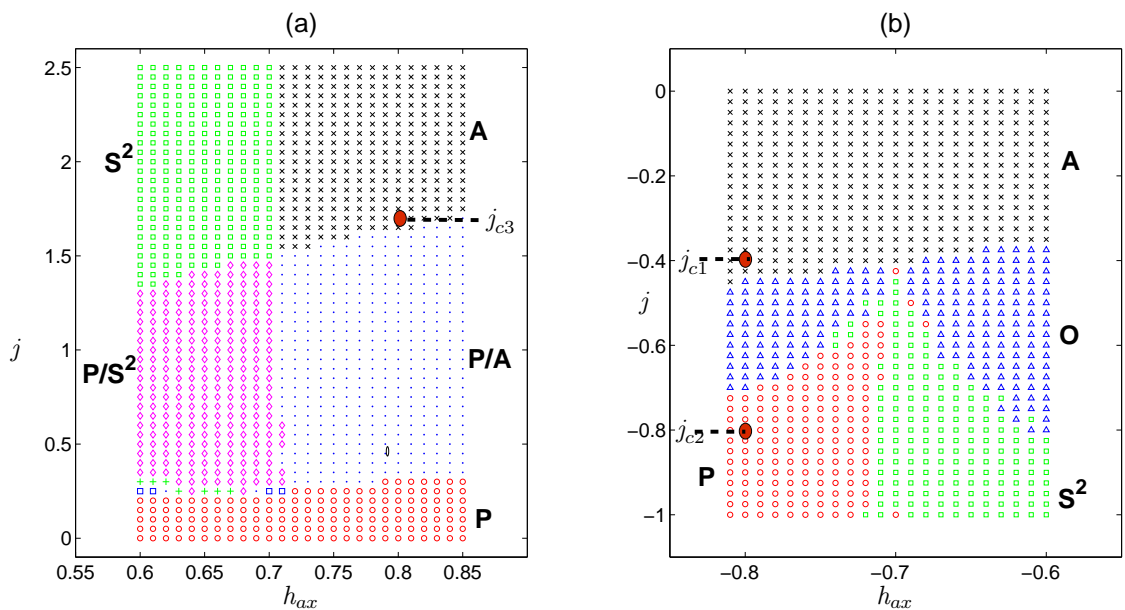}

\caption{Phase diagram in the $h_{ax}-j$ space, in regions relevant for the (a) NOT and (b) AND and OR gates. The system displays limit cycles(O), symmetric out-of-plane limit cycles\,(O$^2$), stable fixed points parallel to $\hat{x}$(P) or $-\hat{x}$(A), and symmetric out-of-plane stable fixed points (S$^2$). The critical value of the current and the field used for our models ($j_{c1},j_{c2}$ and $j_{c3}$) are circled in the two figures.}

\end{figure}
\end{flushleft}

For our choice of geometry, described in the previous section, the magnetization in the free layer exhibits a variety of dynamics 
in different regions of the $h_{ax}-j\,(\equiv\beta/\alpha)$ parameter space - such as in-plane limit cycles (O), symmetric out-of-plane limit cycles (O$^2$), stable fixed points parallel to $\hat{x}$(P), parallel to $-\hat{x}$(A) and symmetric out-of-plane stable fixed points (S$^2$)\cite{bert:2005,bert:2008}. In Fig. 1 we show two specific ranges where the models we propose can perform the desired logical operations. The type of dynamics in the different regions of the parameter space is identified here by 
numerically simulating the LLG equation \ref{llg}. These results clearly agree with those obtained analytically in \cite{bert:2005,bert:2008}. We choose the system parameters $\alpha=0.01$, $D_x=-0.034$, $D_y=0$, and $D_z=0.68$ (as in ref. \cite{kiselev:2003}). Taking the value of saturation magnetization, $M_s$, to be that of Co ($1.4\times~10^6$~A/m), it effectively implies a time scale of $3.2$~ps. 

Magnetic tunnel junctions (MTJs)  have proved themselves to be more worthwhile candidates as MRAMs, with their operability at much lower spin-current and field amplitudes, and higher ferromagnetic to anti-ferromagnetic current ratios\cite{alan:2009,parkin:2003,daughton:1997}. Although the STT phenomenon in MTJs and that in spin-valve pillars display several qualitative similarities, MTJs are hamperedby the lack of an appropriate mathematical model to describe their dynamics. 
We believe results presented in this paper will be of relevance in MTJs too and may possibly be reproduced. Our numerical simulations show that the model presented is robust with respect to errors that may creep in through two of the system parameters - variations in the degree of polarization, and in plane anisotropy fields in the form of $D_x$. We have varied these values 
upto 10\% and yet noticed no percievable difference in the phase diagram. 
A schematic diagram of the spin-valve assembly, with current as input to a two input magneto-logic gate, is shown in the Fig. 2. 

\begin{figure}[h]\label{schematic}
\begin{center}
\includegraphics[width=0.4\linewidth]{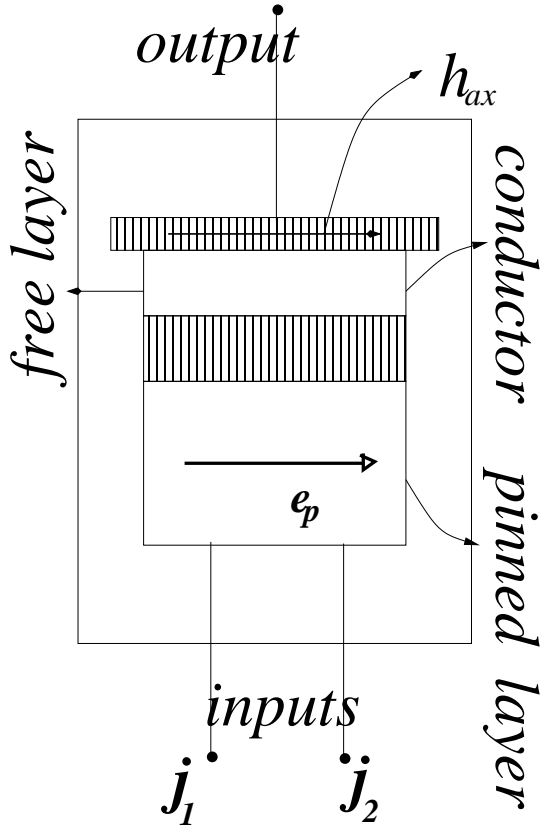}
\caption{A schematic representation of the spin valve assembly used as a two input, universal or non-universal, magneto-logic gate.}
\end{center}
\end{figure}

\section{Logic AND Gate}

\begin{figure}[h]\label{fp}
\begin{center}
\includegraphics[width=0.75\linewidth]{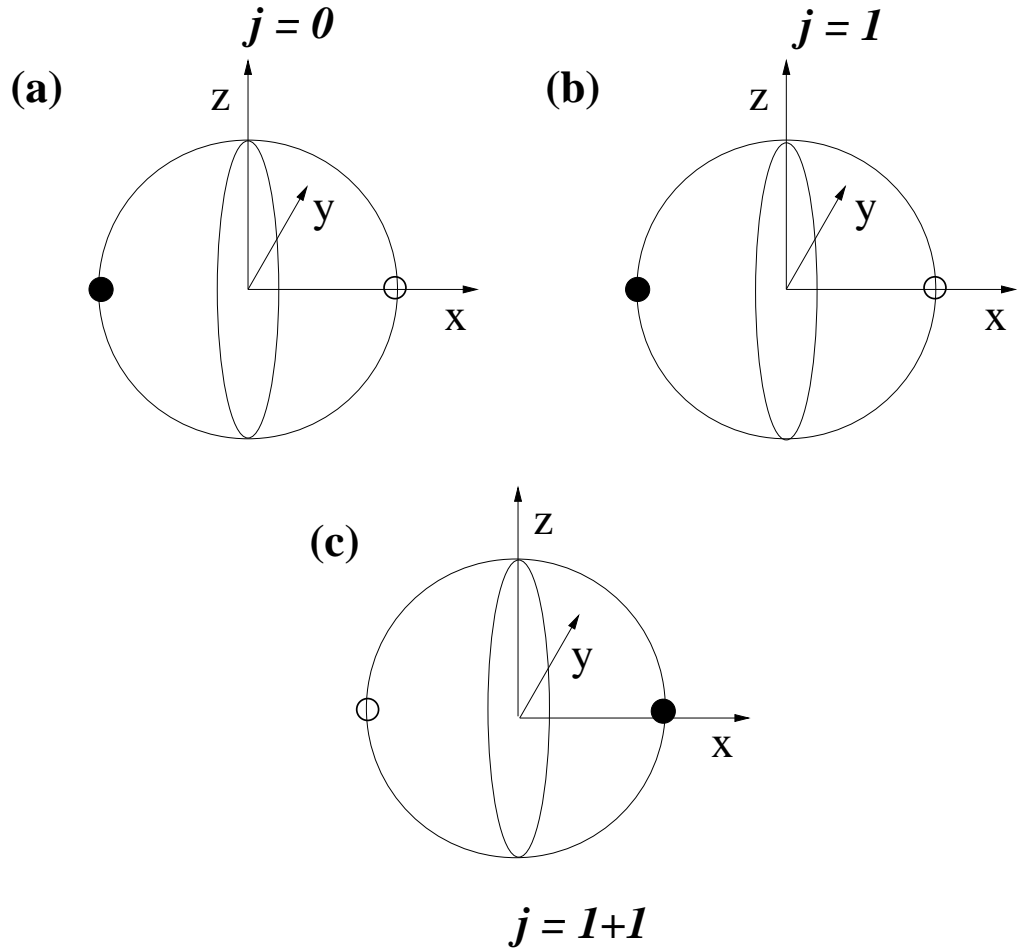}
\caption{Fixed points for three different values of the current $j$, (a) $j=0$, (b) $j=1\,(j_{c1})$ and (c) $j=1+1\,(2j_{c1})$. The applied field $h_{ax}$ is the same, and is negative with $|h_{ax}|>D_z-D_x$ which is represented by $h_{ax}=$1. Both the currents should be 'high' in order to have a 'high' output. }
\end{center}
\end{figure}

For the logical AND gate design consider the fixed points corresponding to a specific negative region of $h_{ax} - j$ parameter space [Fig. 1(b)]. The applied field $h_{ax}$ is chosen to be {\it negative} (whenever non-zero) and and $|h_{ax}|>D_z-D_x$. For this choice, there can at best be only one stable fixed point, lying along either $\pm\hat{x}$ directions  depending on the values of $h_{ax}$ and $j$. The spin current $j$ assumes either of the three values, {\it zero}, $j_{c1}$ or  $2.0j_{c1}$ [where $j_{c1}$ is indicated in Fig. 1(b)]. The currents used in the simulations are atleast 1\% higher than the critical currents to ensure the robustness of the device against random noises. Notice that $j_{c1}$ is {\it negative}, implying a current sent in the opposite direction along the pillar.  The fixed points corresponding to different combinations of $h_{ax}$ and $j$ are shown in Fig. 3. We shall denote the above mentioned negative value of the magnetic field as $h_{ax} =-1$. Also we shall take the current value $j=0$, and $j=j_{c1}$ as the logical inputs 0 and  1, respectively. In the absence of both current and field, the stable fixed points are $\pm\hat{x}$. When the field $h_{ax}=-1$ and the current is either 0 or 1, $\hat{m}=-\hat{x}$ is the only stable fixed point while $\hat{m}=\hat{x}$ becomes unstable. When the current value $j=2.0j_{c1}$ (but same applied field), however, the situation reverses, with $\hat{x}$ becoming stable, and $-\hat{x}$ unstable.

\begin{figure}[h]\label{logic}
\begin{center}
\includegraphics[width=0.75\linewidth]{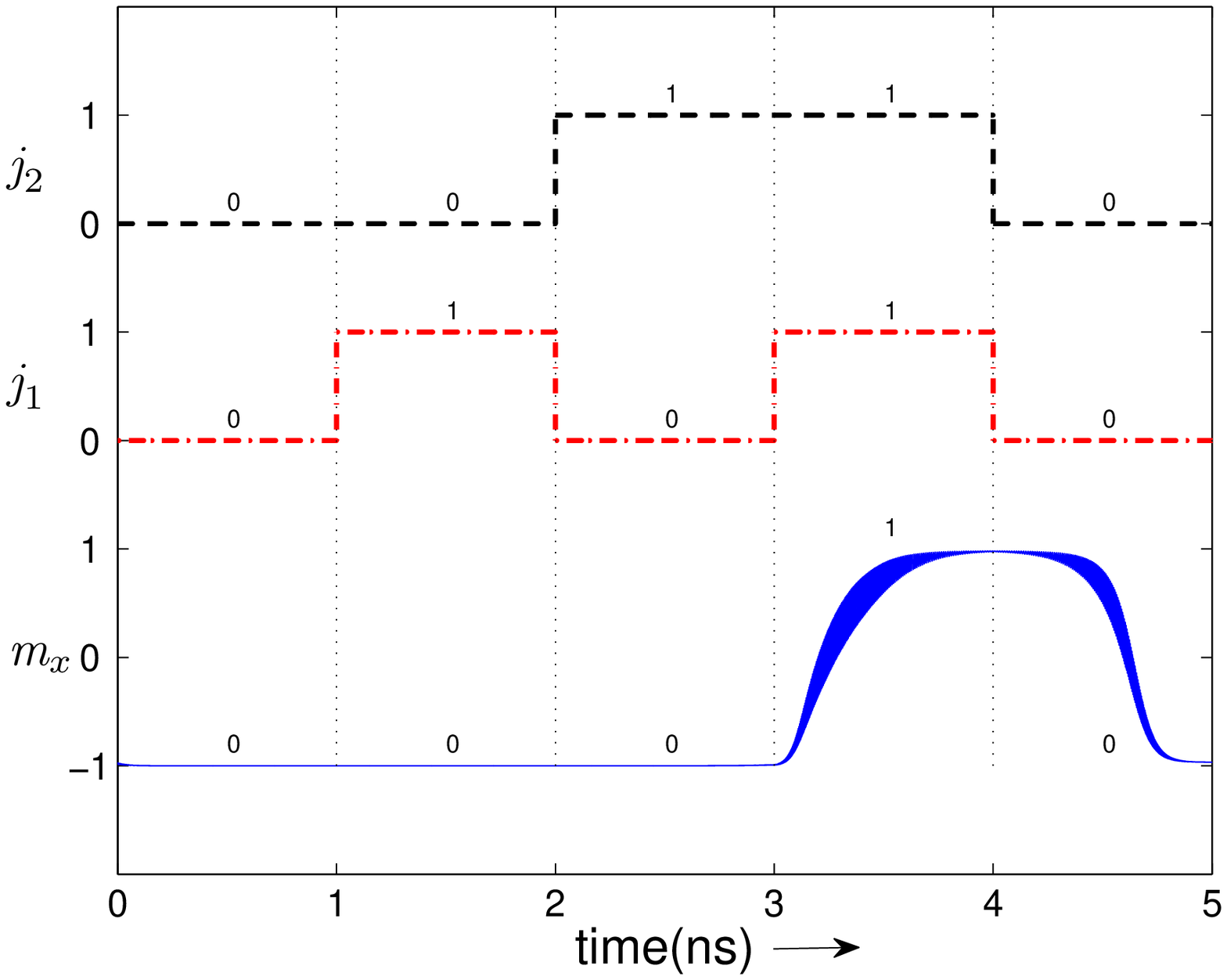}
\caption{Time evolution of $m_x$ (bottom) as the applied currents $j_{1}$ (middle) and $j_{2}$ (top) are flipped through various combinations, relevant to the AND gate. The interpreted logical state is indicated over the respective $m_x$ values. For the parameter values chosen, the switching time is within 1~ns. }
\end{center}
\end{figure}

Let $j_1$ and $j_2$ be the currents fed together, and each take values 0 or 1. The magnetic field is held constant at $h_{ax}=-1$ through out the logical operation. When the free layer magnetization is parallel to the direction of pinning magnetization, we have a low resistance state due to the GMR effect and vice versa. We interpret the high-resistance state ($\hat{m} = -\hat{x}$) as the logical state 0, and the low-resistance state ($\hat{m} = \hat{x}$) as the logical state 1 (The consistency of this convention is preserved in all the gates). The following truth table of the AND operation is thus realized (Table 1). As both $\pm\hat{x}$ are stable fixed points in the absence of current and magnetic field, non-volatility is ensured. Fig. 4 shows the expected response of the free layer magnetization to the flippings of input current $j_{i}$'s.

\begin{table}[h]\label{truth1}
  \begin{center}
    \begin{tabular*}{0.75\columnwidth}{@{\extracolsep{\fill}} | c | c | c | c |}
      \hline
      $h_{ax}$ & $j_{1}$ & $j_{2}$ & ${\bf m}$~(logical state)\\ 
      \hline
      -1 & 0 & 0 & $-\hat{x}$ (0)\\ 
      \hline
      -1 & 1 & 0 & $-\hat{x}$ (0)\\ 
      \hline
      -1 & 0 & 1 & $-\hat{x}$ (0)\\ 
      \hline
      -1 & 1 & 1 & $\hat{x}$ (1)\\ 
      \hline
    \end{tabular*}
    \caption{The truth table for AND gate. The applied field is always held constant through out the operation ($|h_{ax}|>D_z-D_x$) 
      indicated by $h_{ax}=-1$. The currents $j_{1,2}$ take either a value greater than $j_{c1}$, indicated as the logical input 1, or zero taken as input 0.}
  \end{center}
\end{table}


\section{Logic OR Gate}

\begin{figure}[h]\label{fp2}
\begin{center}
\includegraphics[width=0.75\linewidth]{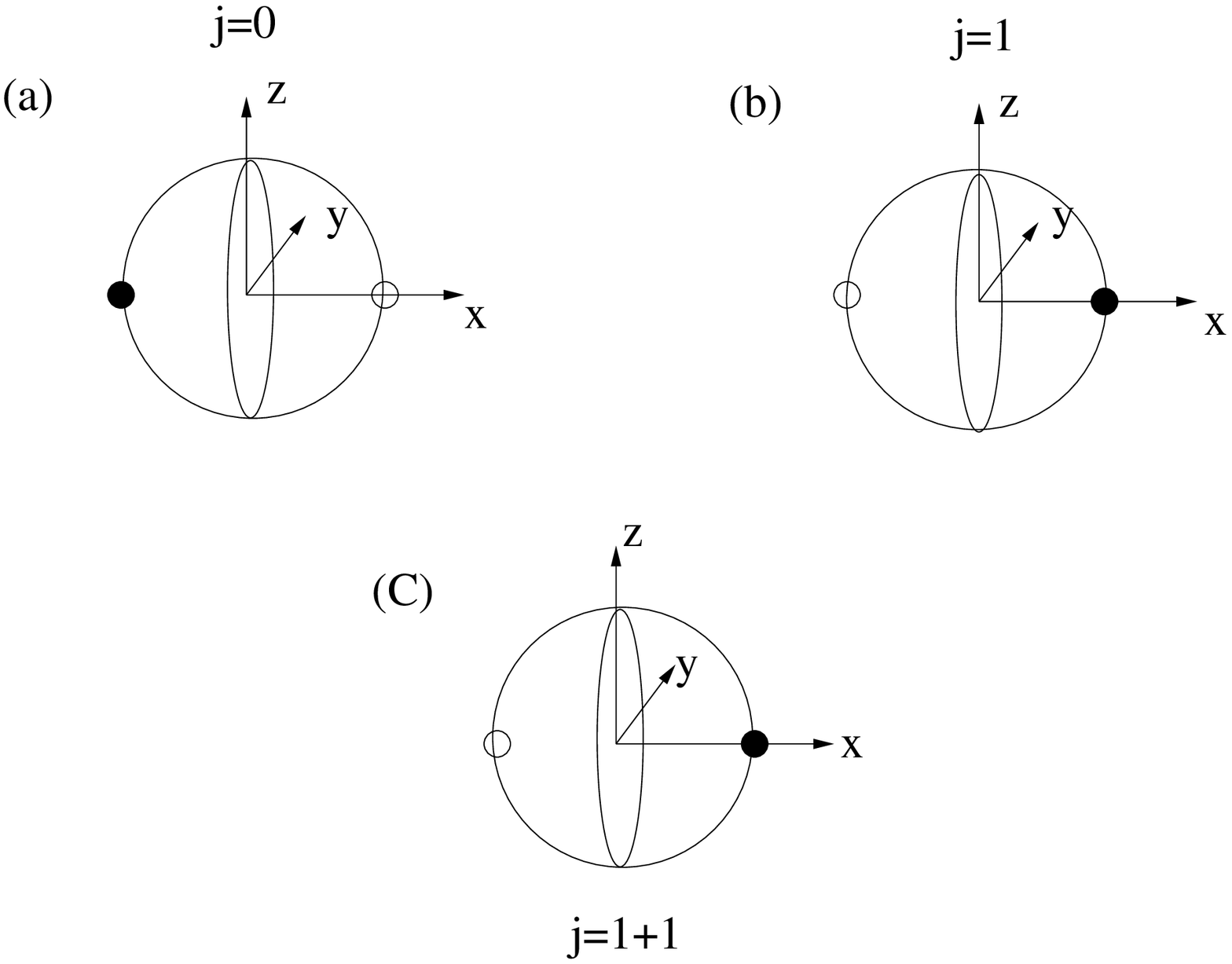}
\caption{Fixed points for three different values of the current $j$, (a) $j=0$, (b) $j=1\,(j_{c2})$ and (c) $j=1+1\,(2j_{c2})$. The applied field is the same as in Fig. 3. It is seen that a single 'high' input current drives the output to a 'high' state. }
\end{center}
\end{figure}

We again consider the region of parameter space we used for logic AND gate [Fig. 1(b)], but we now identify another critical current density value denoted by $j_{c2}$ which can be used for implementing the logic OR gate. Agian the applied field $h_{ax}$ is chosen to be {\it negative} (whenever non-zero) and and $|h_{ax}|>D_z-D_x$. Now the spin current $j$ assumes either of the three values, {\it zero}, $j_{c2}$ or  $2.0j_{c2}$ [Again, $j_{c2}$ is indicated in Fig. 1(b)]. The fixed points corresponding to different $j$ values are shown in Fig. 5. We shall take the current value $j=0$, and $j=j_{c2}$ as the logical inputs 0 and  1, respectively. In the absence of both current and field, the stable fixed points are $\pm\hat{x}$. When the field $h_{ax}=-1$ and the current $j_{i}=$0, $\hat{m}=-\hat{x}$ is the only stable fixed point while $\hat{m}=\hat{x}$ becomes unstable. When either of the input current value $j_{i}=$1 ($j_{i}=1.0j_{c2}, i=1,2$), the situation reverses, with $\hat{x}$ becoming stable, and $-\hat{x}$ unstable. The same scenario repeats with the current value $j=$1+1 ($j=2.0j_{c2}$) realizing a magneto-logic OR gate. 

\begin{figure}[h]\label{logic}
\begin{center}
\includegraphics[width=0.75\linewidth]{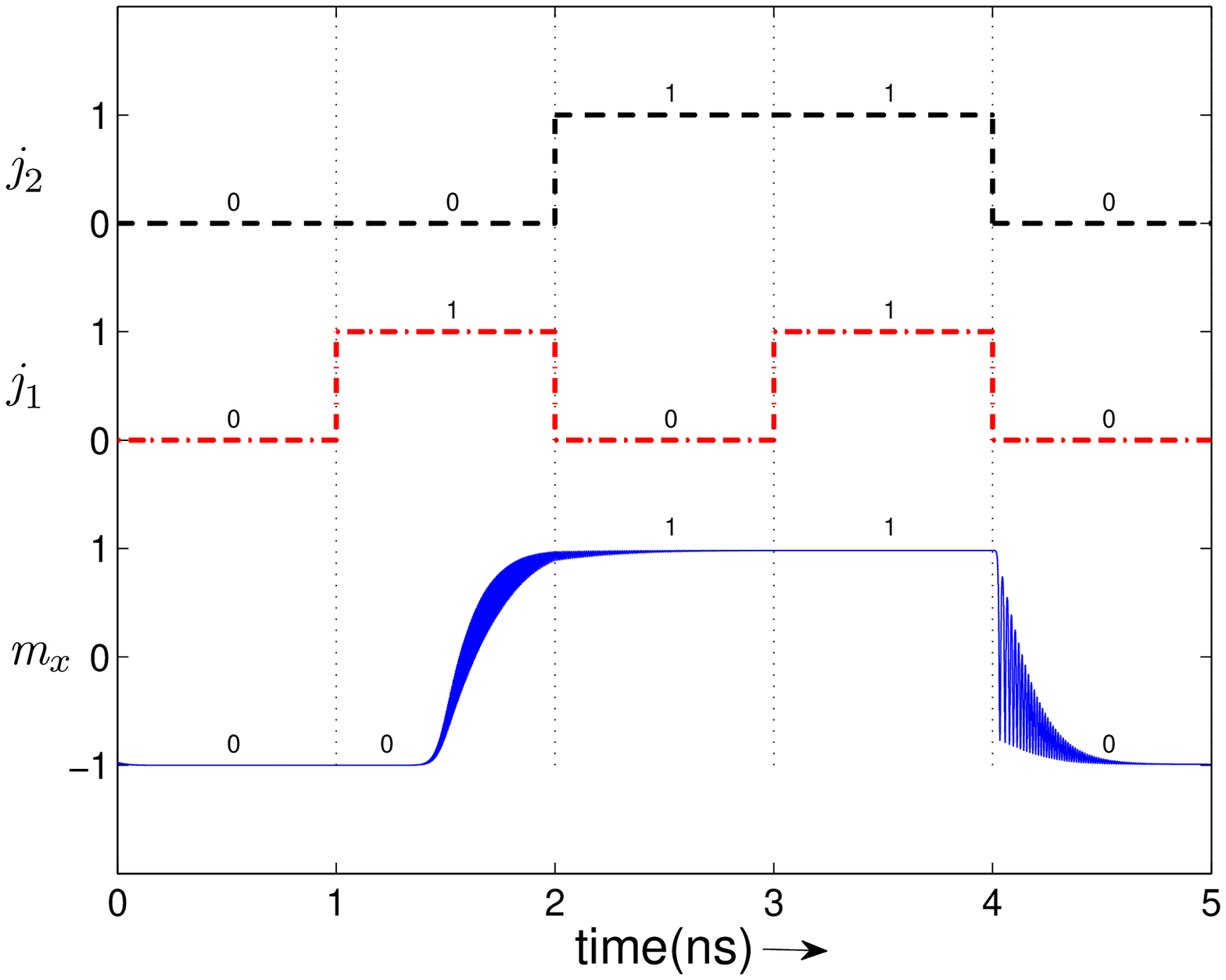}
\caption{Time evolution of $m_x$ (bottom) as the applied currents $j_{1}$ (middle) and $j_{2}$ (top) are flipped through various combinations, relevant to the OR gate. The interpreted logical state is indicated over the respective $m_x$ values. For the parameter values chosen, the switching time is within 1~ns. }
\end{center}
\end{figure}

The truth table of magneto-logic OR gate is shown in the table 2, and numerical results for the flipping of $x$-component of free layer magnetization is shown in Fig. 6.

\begin{table}[h]\label{truth2}
\begin{center}
	\begin{tabular*}{0.75\columnwidth}{@{\extracolsep{\fill}} | c | c | c | c |}
	\hline
	$h_{ax}$ & $j_{1}$ & $j_{2}$ & ${\bf m}$~(logical state)\\ 
	\hline
	-1 & 0 & 0 & $-\hat{x}$ (0)\\ 
	\hline
	-1 & 1 & 0 & $-\hat{x}$ (1)\\ 
	\hline
	-1 & 0 & 1 & $-\hat{x}$ (1)\\ 
	\hline
	-1 & 1 & 1 & $\hat{x}$ (1)\\ 
	\hline
	\end{tabular*}
\caption{The truth table for OR gate. The applied field is always held constant through out the operation ($|h_{ax}|>D_z-D_x$) 
indicated by $h_{ax}=-1$. The currents $j_{1,2}$ take either a value greater than $j_{c2}$, indicated as the logical input 1, or zero taken as input 0.}
\end{center}
\end{table}

\section{Logic NOT gate}

\begin{figure}[h]\label{fp3}
\begin{center}
\includegraphics[width=0.75\linewidth]{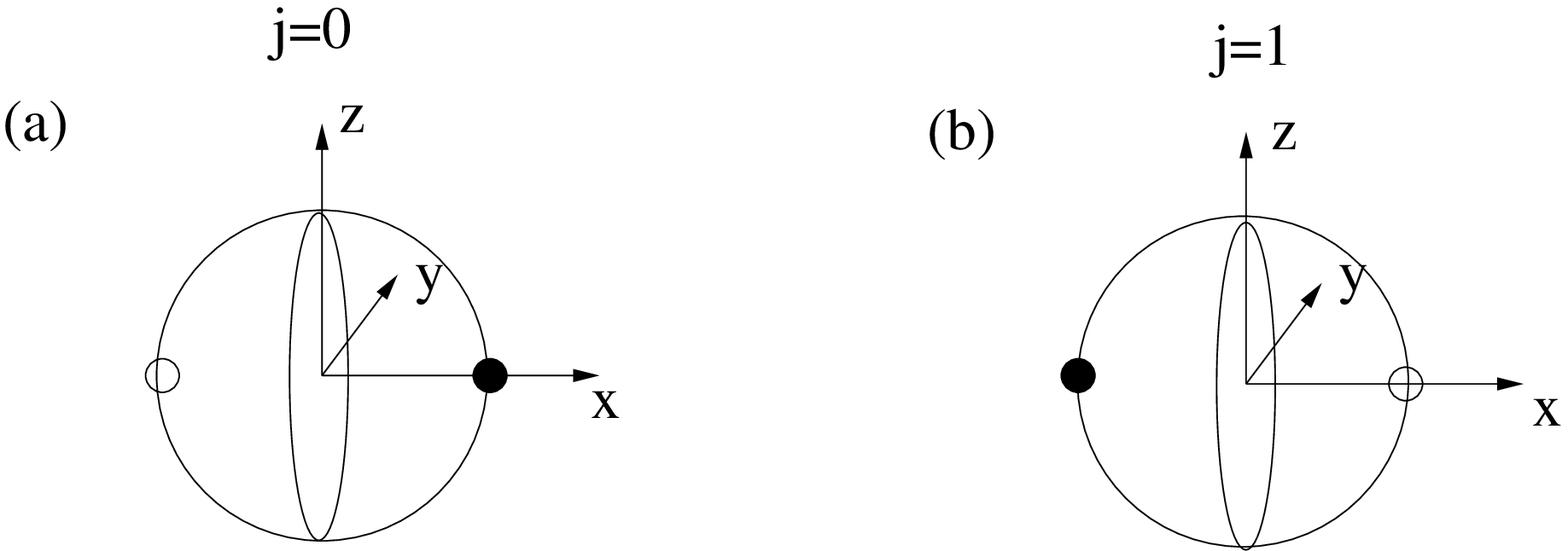}
\caption{Fixed point illustration for NOT gate. (a) $j=0$ and (b) $j=1\,(j_{c3})$. The applied field is the same as Fig 3 in magnitude but is now pointing in the $+\hat{x}$ direction denoted by $h_{ax}=$1. It is seen that a 'high' input current drives the output to a 'low' state realizing NOT gate. }
\end{center}
\end{figure}

In order to realize the logic NOT gate, we turn our attention to the selected positive side of the $h_{ax}-j$ control space shown in Fig 1. (a). Now $h_{ax}$ is held at the same numerical value as earlier but now in the $+\hat{x}$ direction denoting $h_{ax}=1$ in the following discussion. We use the critical current density denoted by $j_{c3}$ in the Fig 1. (a) to realize the NOT gate. The spin current density $j > j_{c3}$, but within the same dynamic regime in phase space, is denoted by $j=$1. From the nature of fixed points illustrated in Fig. 7, it is clear that whenever the current density toggles from logical values $j=0$ to $j=1$, the nature of stability interchanges between the two fixed points $\hat{x}$ and $-\hat{x}$, and vice-versa. With our interpretation of $+\hat{x}$ as logical state 1 and $-\hat{x}$ as state 0, we have an immediate realization of NOT gate (See Table 3. for the truth table).

\begin{table}[h]\label{truth3}
  \begin{center}
    \begin{tabular*}{0.75\columnwidth}{@{\extracolsep{\fill}} | c | c | c |}
      \hline
      $h_{ax}$ & $j$ & ${\bf m}$~(logical state)\\ 
      \hline
      1 & 0 & $+\hat{x}$ (1)\\ 
      \hline
      1 & 1 & $-\hat{x}$ (0)\\ 
      \hline
    \end{tabular*}
    \caption{The truth table for NOT gate. The applied field is once again held constant through out the operation ($|h_{ax}|>D_z-D_x$), 
      indicated by $h_{ax}=1$. The current $j$ take either a value greater than $j_{c3}$, indicated as the logical input 1, or zero taken as input 0.}
  \end{center}
\end{table}

\section{Summary}
In summary, we have proposed spin-valve based magneto-logic AND, OR and NOT gate assemblies, which render themselves to the dual role of universal gate and a magnetic memory.  An applied magnetic field, held constant through out,  parallel to the pinned layer magnetization acts as a control for the logic gate operation, while spin-currents are fed in as the logical inputs. The same pillar geometry is used for all the proposed gates, which also doubles as a magnetic memory device. The gates are programmable and its functionality can be changed by the direction of control field and current. The critical current densities, used to represent logical input, are differ for each type of gate. There is, however, consistency in the interpretation of the logical inputs and outputs, with respect to the direction, or sign, of the current densities.

\end{document}